\documentstyle[amstex,graphics,epsfig,amssymb,preprint,prd,aps]{revtex}
\begin{document}
\tightenlines
\draft
\title{Quintessence or Phoenix?}
\author{Claudio Rubano}
\address{Dipartimento di Scienze Fisiche, Universit\`{a} Federico II\\
and \\ Istituto Nazionale di Fisica Nucleare, Sez. di Napoli,\\ Complesso
Universitario di Monte S. Angelo,\\ Via Cinthia, Ed. G, I-80126 Napoli,
Italy\\ }
\author{Paolo Scudellaro}
\address{Dipartimento di Scienze Fisiche, Universit\`{a} Federico II\\
and \\ Istituto Nazionale di Fisica Nucleare, Sez. di Napoli,\\ Complesso
Universitario di Monte S. Angelo,\\ Via Cinthia, Ed. G, I-80126 Napoli,
Italy}
\date{\today}
\maketitle

\begin{abstract}
We show that it is impossible to determine the state equation of
quintessence models on the basis of pure observational SNIa data. An
independent estimate of $\Omega_{M0}$ is necessary. Also in this most
favourable case the situation can be problematic.

PACS number(s): 98.80.Cq, 98.80.Hw, 04.20.Jb
\end{abstract}

\narrowtext

\section{Introduction}

\smallskip

\emph{\indent\indent... l'araba fenice}

\emph{Che vi sia, ciascun lo dice}

\emph{Dove sia, nessun lo sa}\footnote{''...
Everybody says//The phoenix is there,//But no one knows where'', in
``Cos\`{\i} fan tutte'', by L. da Ponte.}

\bigskip

In the history of cosmology there are many cases of Arabic phoenixes.
The metaphor applies particularly well to the cosmological constant,
which seems to resurrect out of its ashes and challenges any
interpretation since almost one century \cite{Carroll,Sahni}. In its
last resurrection (that is, quintessence
\cite{Ostrik,CDS,ZWS,SWZ,Wang}), it poses the problems of determining
the state equation of the peculiar quintessential fluid and/or  the
right potential of the associated scalar field.

In this paper we want to investigate some subtleties in the socalled
`reconstruction of the state equation' on the basis of observational data,
showing that it is probably impossible to determine the state equation and
even the value of $\Omega_{M0}$ only on the basis of pure observations. In
other words: you cannot find the phoenix if you have no idea about its
face.

The starting idea came from a paper by A.A. Sen and S. Sethi
\cite{Sen}, in which they propose an ansatz for the Hubble parameter
as a function of $t$ and a suitable parameter $\beta$. They find that
it is possible to obtain a good agreement with present data on SNIa
for a certain range of values of $\beta$. A particular choice allows
then the exact evaluation of the scalar field potential. The surprise
is that the value of $\Omega_{M0}$ is obtained in terms of an
integration constant. This means that the theory fits the data for any
arbitrary value of this parameter!

Another result in this direction is due to I. Wasserman \cite{Wass}.
He finds that the usual expression for $H(z)$, in the case of presence
of a $\Lambda$-term plus dust, can be analitically derived from a
quintessence potential with an independent choice for the value of
${\Omega}_{M0}$.

In  Sec. 2 of this paper we generalize these results, showing that
they do not depend on the particular ansatz, so that any empiric
evaluation of the state equation must be supplied with a value for
$\Omega_{M0}$, obtained by independent observations.

In Sec. 3 we show that also in the ideal situation of almost infinite
precision in observational data, the reconstruction of the state equation
could be impossible.

In Sec. 4 we consider three exactly integrable models, one found by us
\cite{Rub}, and those studied by Sen and Sethi \cite{Sen} and
Wasserman \cite{Wass}, showing that they all can perfectly mimic a
$\Lambda$-term model.

In Sec. 5 discussion and conclusions are given.

\section{Reconstruction of the state equation}

Let us  consider a spatially flat universe, minimally coupled with a
scalar field, and adopt the conventions $8\pi G=1$, $c=1$, $a_{0}=1$,
$H_{0}=1$, where $a_{0}$ is the present day scale factor, and $H_{0}$
is the value of the Hubble constant (in appropriate units). This can
be done without any loss of generality.

Starting from the Friedman and Klein-Gordon equations
\begin{gather}
3H^{2}=\frac{1}{2}\dot{\varphi}^{2}+V(\varphi)+3\Omega_{M0}
a^{-3}\label{eq1}\,,\\
\ddot{\varphi}+3H\dot{\varphi}+ \frac{dV}{d\varphi}=0\,,\label{eq2}
\end{gather}
it is possible (differentiating Eq.(\ref{eq1}) and with easy algebraic
manipulation) to derive
\begin{align}
\dot{\varphi}^{2} &  =-2\dot{H}-3\Omega_{M0} a^{-3}\label{eq3}\,,\\
\varphi &  =\int\sqrt{-2\dot{H}-3\Omega_{M0} a^{-3}}dt\label{eq4}\,,\\
V &  =3H^{2}+\dot{H}-\frac{3}{2}\Omega_{M0} a^{-3}\,.\label{eq5}
\end{align}

The last two equations give a parametric expression for $V(\varphi)$,
once an ansatz for $a(t)$, or $H(t)$, is given. The point is that also
a value for $\Omega_{M0}$ should be supplied.

The situation is complicated by the fact that the ansatz contains parameters
whose link with observations could be rather difficult to find, if not
impossible. Thus, we prefer to illustrate a  situation nearer to the
observational strategy.

Assume that it is possible to observe an enormous number of type Ia
supernovae with extreme precision, so that it is possible to construct
an empirical function for the luminosity distance versus the redshift
(this is in fact what is really measured in the supernovae
experiments), say $d_{L}(z)$. \emph{It will contain some numerical
coefficients whose physical interpretation could be difficult or even
impossible.} (In the case of time dependence, on the contrary, the
physical meaning of the parameters is usually clear.)

From $d_{L}(z)$ it is possible to derive $H(z)$. With our
normalization, the definition of $d_{L}(z)$ is
\begin{equation}
d_{L}(z)=(1+z)\int_{0}^{z}\frac{dz^{\prime}}{H(z^{\prime})}\,,\label{eq6}
\end{equation}
which gives
\begin{equation}
H(z)=\frac{(1+z)^{2}}{(1+z)d_{L}^{\prime}(z)-d_{L}(z)}\,.
\end{equation}

Here and below, prime stands for derivation with respect to $z$. From
Eqs.(4) and (5), it is now possible to write
\begin{align}
\varphi(z) &  =\int\sqrt{\frac{2H^{\prime}}{H(1+z)}-\frac{3\Omega_{M0}(1+z)}{H^{2}}
}\label{eq8}\,,\\ V(z) &
=3H^{2}-HH^{\prime}(1+z)-\frac{3}{2}\Omega_{M0}(1+z)^{3}\,,\label{eq9}
\end{align}
which now allow to obtain $V(\varphi)$ starting from $d_L (z)$ (we
give an example below). But the most interesting result is the
expression of the state equation versus the redshift
\begin{equation}
w=\frac{2HH^{\prime}(1+z)-3H^{2}}{3H^{2}-3\Omega_{M0}(1+z)^{3}}\,.\label{eq10}
\end{equation}

It is then clear that the value of $\Omega_{M0}$ must be supplied
independently, and that the model obtained works perfectly with any
value!

\section{But things stay even worse!}

The situation is, however, complicated by other subtleties. Let us illustrate
it with an example, which should also make clearer the previous arguments.

Let us take as fiducial model a simple $\Lambda$-term model, with
$\Omega_{M0}=0.3$ (which is of course irrelevant and is taken in
homage to current fashion). It is then easy to produce a data set of
200 points for $d_{L}(z)$ , in the range $z=0\div2$. The precision is
that of numerical integration of the MATHEMATICA algorithm, i.e.,
practically infinite. Assume also that $\Omega_{M0}$ has been
determined independently, again with almost infinite precision.

Now we find an empirical $d_{L}(z)$ as a quartic polynomial, by means
of a best fit. Having set $H_{0}=1$, we keep fixed the first
coefficient and get
\begin{equation}
d_{Lfit}(z)=z+0.760z^{2}-0.257z^{3}+0.040z^{4}\,.\label{eq11}
\end{equation}
The fractional difference with the data set is $<0.002$.

If we find $w(z)$ according to Eq. (\ref{eq11}), and with the correct
value $\Omega_{M0}=0.3$, we obtain the plot in Fig. 1. It is clear at
a first glance that $w$ is far from being constant, but the real
problem is that the values $w < -1$ are absurd (in this context).
Indeed, they imply ${\dot{\varphi}}^2 < 0$ and are of course an
artifact of the procedure; this could be interpreted as a signal of
something being wrong.

But we have also to remember that the value ${\Omega}_{M0}=0.3$ is supposed
to come from independent observations. We know that the precision of such
observations is presently very low, and we can figure out that, also in the
ideal situation here examined, it is well possible to consider a slightly
different value of $0.28$. In this case we have no problems up to $z \sim
1.7$ (see Fig. 2), and it is possible to `reconstruct' a potential like
that in Fig. 3, which looks nice but, of course, has nothing to do with the
starting point of our analysis. The situation is not substantially changed
even if we go up to a $7$th-degree polinomial as best fit.

The last desperate trial is to use, instead of the best fit, an
interpolation of the data set. In this case all is made with the
internal precision of MATHEMATICA (16 digits) and gives for $w$ the
plot in Fig. 4, and only at this level it is clear that the variations
are due to numerical computation.

\section{May a portrait of the phoenix be of help?}

A possible objection to the argument of Sec. 3 is that an empirical
polynomial is a very crude assumption, and that we should try with specific
models. In this case the infinite amount of possibilities poses some
problems of choice. We present here three possible models, which have the
nice property of being exact solution of the equations, so that all the
considerations are very clear and no approximation error can be invoked.

The first model is given by a potential already studied  by us
\cite{Rub,Pavlov,Sereno}(but see also \cite{Card}), which shows a simple
exponential dependence on $\varphi$
\begin{equation}
V=V_{0}\exp(-\sqrt{\frac{3}{2}}\varphi)\,.
\end{equation}

For the details of the procedure for finding the solution and for the
subsequent discussion on its properties, see \cite{Rub}. Here we limit
ourselves to present the expression of $H(t)$ and $z(t)$, adapted to
our normalizations
\begin{align}
H(t)  &
=\frac{(1+2t^{2})(t_{0}+t_{0}^{3})}{t(1+t^{2})(1+2t_{0}^{2})}\,,\label{eq13}\\
z(t) &
=\sqrt[3]{\frac{t^{2}(1+t^{2})}{t_{0}^{2}(1+t_{0}^{2})}}-1\,.\label{eq14}
\end{align}

The variable $t$ could be eliminated, in order to have $H(z)$ explicitly,
but there is no need of doing this, and everything can be made by treating
Eqs. (\ref{eq13}) and (\ref{eq14}) as defining a parametric dependence. The
only parameter is $t_{0}$, which is linked to the value of $\Omega_{M0}$,
by
\begin{equation}
\Omega_{M0}=\frac{1+t_{0}^{2}}{(1+2t_{0}^{2})^{2}}\,.
\end{equation}

This model has been tested \cite{Rub,Pavlov,Sereno} with the currently
available data on SNIa \cite{Perl1,Perl2,Riess,Garn,Jha} and on peculiar
velocities \cite{Pea}, and seems to indicate a value for $\Omega_{M0}$ much
lower than the usually indicated one, but there is no definite evidence for
this. In any case a range like $\Omega_{M0}=0.15\div0.30$, is fully
compatible.

We now compare the results of this model with the fiducial
$\Lambda$-term, in the \emph{ideal} case. It should be clear that
there is no reason why the value of $\Omega_{M0}$ should be the same
in the two cases: it is a free parameter of the theory, which has to
be adapted to data, and the results can be very different. If we set
$t_{0}=1$, corresponding to $\Omega_{M0}=0.22$, and compare
$d_{L}(z)$, from Eqs.(\ref{eq13}) and (\ref{eq14}) with the above
fiducial model, we obtain for the fractional difference an agreement
up to $2\%$.

In this case, the value of $\Omega_{M0}$ is an important element of
the model, and the fact that it is significantly different in the two
cases means that an independent measure would be of great help in
distinguish them. The problem is in the very poor accuracy of this
kind of determination, but we do not want to examine observational
technicalities in this paper.

Let us instead present a case in which the value of $\Omega_{M0}$ is
unpredictable, as announced before. In \cite{Sen} Sen and Sethi
present an ansatz which, adapted to our normalizations, has the form
\begin{equation}
H(z)=\tanh(1)\sqrt{1+\frac{(1+z)^{2/\beta}}{\sinh(1)^{2}}}\,,\label{eq16}
\end{equation}
where $\beta$ is a parameter which has to be adjusted from data. They
find as best fit value for current data  $\beta=0.81$, but
$\beta=2/3$ is still compatible, and has the advantage of leading to
an analytic expression for the potential
\begin{equation}
V(\varphi)=\frac{A^{2}}{8}(\exp(2B\varphi)+\exp(-2B\varphi))+V_{0}\,,
\end{equation}
where $A$ is an arbitrary integration constant and, with our
normalization,
\begin{equation}
B=\frac{3}{2A\coth(1)}\,,\quad V_{0}=3\tanh(1)^{2}-\frac{A^{2}}{4}\,.
\end{equation}

The interesting fact is that $A$ is correlated to the value of
$\Omega_{M0}$ by the relation
\begin{equation}
\Omega_{M0}=\frac{4-3A^{2}}{\sinh(1)^{2}}\,,
\end{equation}
so that any value is allowed, provided that $d_{L}(z)$ derived from
Eq.(\ref{eq16}) fits the data. It is important to note that the
special value of $\beta=2/3$ only gives analytic evaluation, but any
value can be used, leading to a situation similar to that in Sec. 2,
but with a much more reasonable ansatz for $H$. Assuming for
simplicity $\beta=2/3$, we can easily show that this model mimics a
$\Lambda$-term one. In this case, since no free parameter is left, we
have to change the value in the fiducial model. Taking
$\Omega_{\Lambda}=?$, we again obtain a fractional difference less
than $2\%$. We see that also in this case an independent estimate of
$\Omega_{M0}$ would be of help, but only if we stick to a particular
value of $\beta$.

Still more interesting is the situation illustrated by Wasserman in
\cite{Wass}, where the match of a $\Lambda$-term model with a quintessence
model is analitically exact, and yet the value of $\Omega_{M0}$ is
arbitrary. It is also interesting that the potential found in this paper is
of the same type as in Eq. (17).

\section{Conclusions}

As said in the introduction, if you want to catch the phoenix, you
must have an idea of its aspect. The arguments of Sec.4 show that this
could be not enough. Assuming that the `real' situation is the
presence of a $\Lambda $-term, there is an infinite host of
`reasonable' quintessence models, with unpredictable values of
$\Omega_{M0}$. The theoretical reasons for this are well explained by
Maor and colleagues \cite{Maor}.

Is it then  a black cow in a dark night? May be not completely. A
feature which all these models share is that they are all totally
empiric, i.e., skillful guesses, and are based on (and/or fit)
observations without not so many definite ideas of the precise
physical mechanism behind the proposed potential.

In other words, our opinion is that the above arguments are a sort of
vindication of the theory against excessive trust in the observational
results. The literature is full of papers about the wonderful
perspectives opened by the future observations, and for sure they will
be fundamental in the resolution of the problem. But a satisfactory
model can be only one which has roots in fundamental physics and
interfaces with the general cosmological theory.

Another conclusion which we draw is that a precise measure of
$\Omega_{M0}$, independent of SNIa observations, could be of
fundamental help (although probably not conclusive), but we cannot
figure out how this goal could be reached in short time.

Despite the dramatic improvement in the observational data, which we
expect in the future, the correct extension of the cosmological
standard model still seems a very difficult task.

\bigskip

\noindent {\Large FIGURE CAPTIONS:}
\bigskip

Fig. 1. The manifestly absurd `reconstructed state equation' (see
text).

\medskip

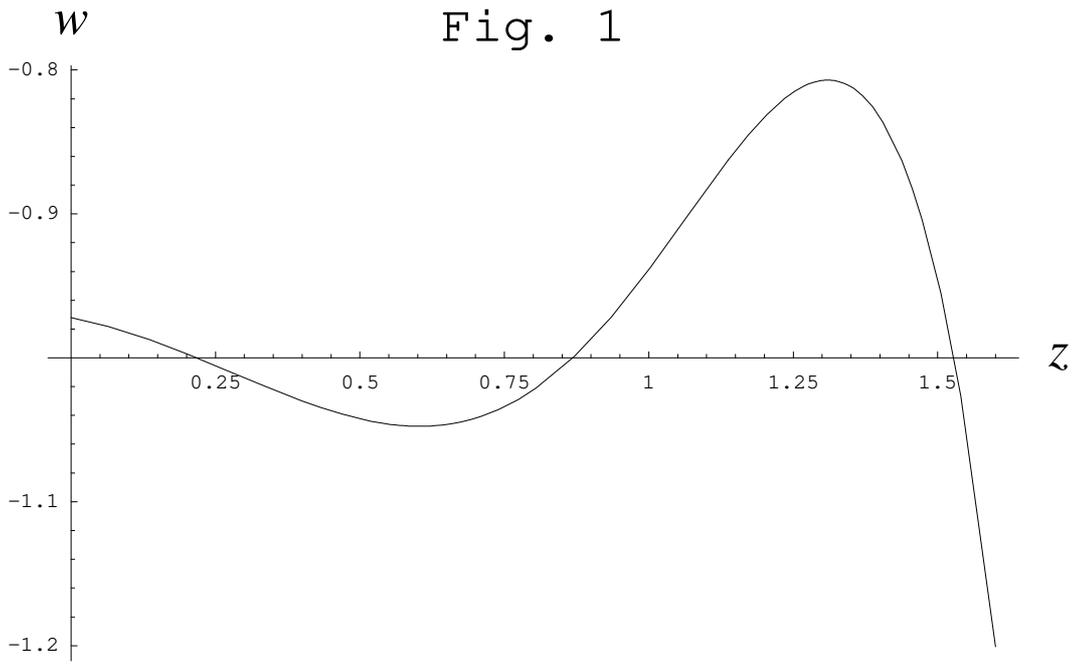
Fig. 2. The state equation with ${\Omega}_{M0}=0.28$. It is still
uncorrect, but only if one knows in advance that it should be $w
= -1$.

\medskip

Fig. 3. The `reconstructed potential', which is of course only an
artifact of the procedure, but seems reasonable.

\medskip

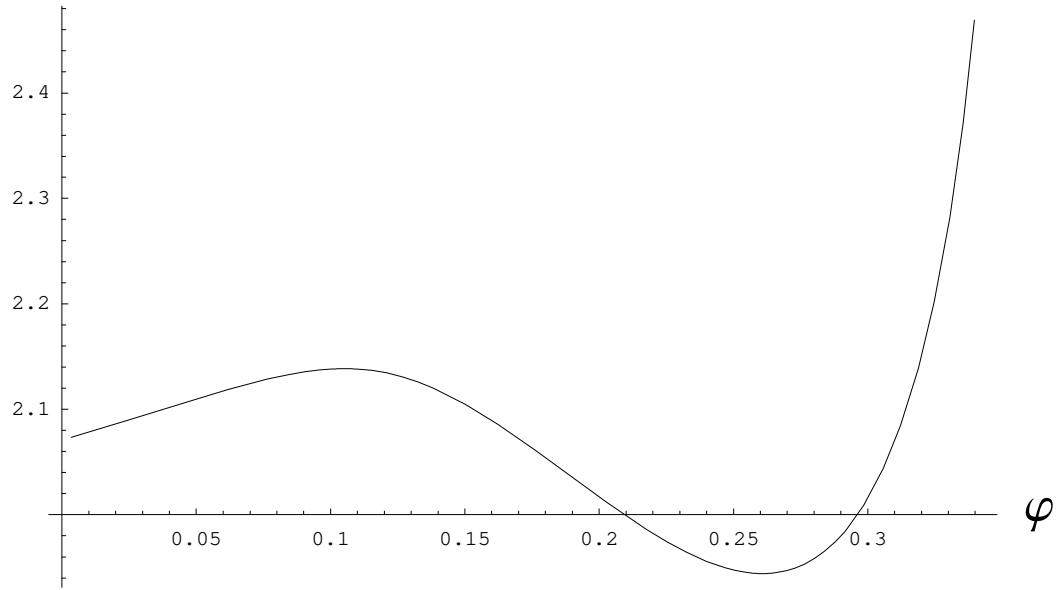
Fig. 4. Eventually, the `correct' result.

\bigskip

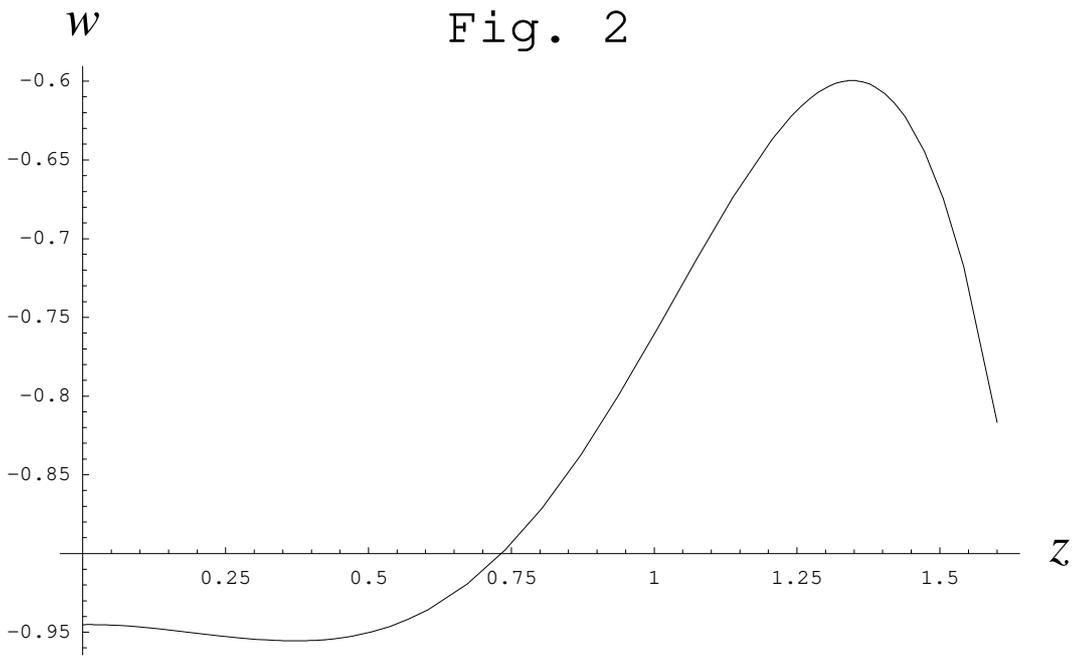
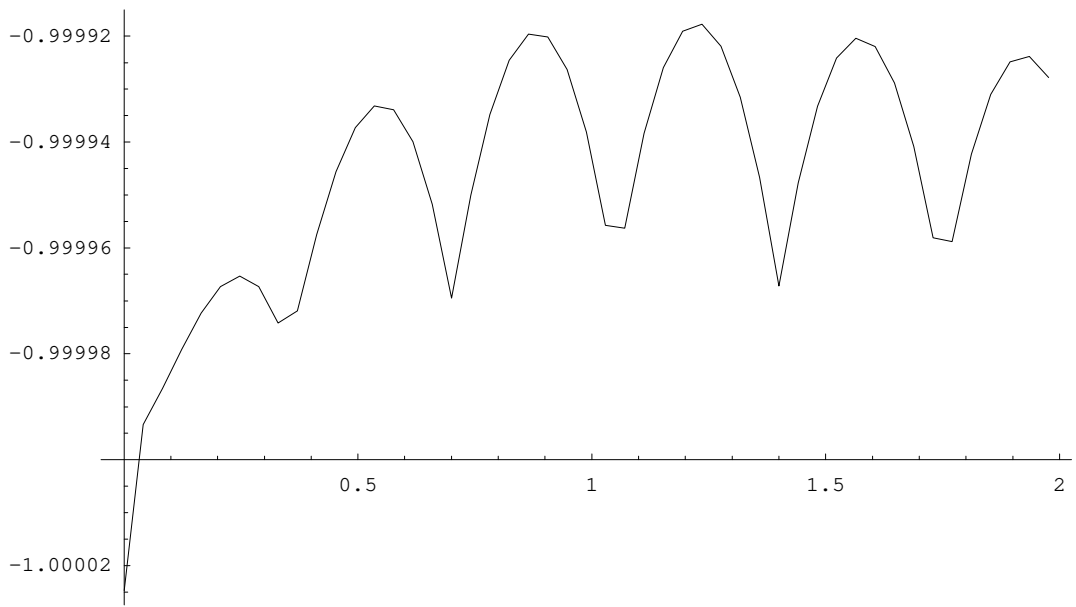
\end{document}